# Scalable massively parallel computing using continuous-time data representation in nanoscale crossbar array

Cong Wang[1,#], Shi-Jun Liang[1,#], Chen-Yu Wang[1], Zai-Zheng Yang[1], Yingmeng Ge[2], Chen Pan[1], Xi Shen[1], Wei Wei[1], Zaichen Zhang[2], Bin Cheng[1], Chuan Zhang[2], Feng Miao[1]*

[1]National Laboratory of Solid State Microstructures, School of Physics, Collaborative Innovation Center of Advanced Microstructures, Nanjing University, Nanjing, China.

[2]National Mobile Communications Research Laboratory, Southeast University; Purple Mountain Laboratories, Nanjing, China.

*Correspondence Email: miao@nju.edu.cn

[#] Equally contribute to this work

**Abstract**

**The switching speed between two discrete logic stages essentially determines the performance of a traditional digital computer based on the von Neumann architecture[1,2]. Unfortunately, the heat wall has prevented the clock speed from increasing since 2004[3]. With the explosive growth of connected intelligent devices in the Internet of Things (IoT), there has been a pressing need for the real-time processing and understanding of the large volume of analogue data generated[4,5]. As a result, the difficulty in boosting the computing speed renders digital computing with a binary data representation unable to meet the increasing demand for processing analogue information that is by nature, intrinsically continuous in magnitude and time[6,7]. By utilizing a continuous data representation, with the inherent nature of physical laws valid at every instant, parallel computing can be implemented for the direct processing of analogue information in real time, without suffering from the constraints of traditional computing hardware[8,9]. Here, we propose a scalable massively parallel computing scheme by exploiting a continuous-time data representation and frequency multiplexing. This computing scheme enables the parallel reading of stored data and the one-shot operation of matrix-matrix multiplications in a memristive crossbar, which are inaccessible via previous computing technologies. Furthermore, we achieve the one-shot**

**recognition of 16 letter images based on two physically interconnected crossbars and demonstrate that the processing and modulation of analogue information can be simultaneously performed in memristive crossbar. Our work paves the way towards the development of intelligent edge devices that are able to process and communicate in real time.**

Discrete- and continuous-time signals are available for information representation in the time domain. A typical example is the binary data representation (i.e., "1" and "0") used for computing in digital computers based on von Neumann architecture, in which the bit stream is discrete in time and amplitude. The upper bound of the processor speed in digital computers is largely set by the clock frequency, the increase of which is essentially limited by the speed of flipping logic states. Further increasing the processor speed will lead to serious overheating issues[1,2], which explains why the clock frequency of advanced digital computers has stopped growing for over ten years[3]. This effect renders digital computing highly challenging in many applications, such as intelligent edge applications in the Internet of Things (IoT) network with the explosive growth of connected edge devices, which require high-efficiency data processing and communication[8-10].

Alternative computing schemes other than digital computing are thus required for these applications[5,11-23]. Different from the discrete data representation, the use of a continuous-time data representation can avoid flips between different logic states and can overcome the aforementioned speed bottleneck of the processor[24]. Processing information represented with a continuous-time signal demands a hardware architecture on which computation can be implemented in a continuous-time manner. Memristive crossbars offer an ideal platform that can implement analogue computing[25-34], in which energy-efficient visual/speech processing and recognition have been achieved[35-39].

In this letter, we propose and implement a scalable massively parallel computing scheme by employing a continuous-time data representation and frequency multiplexing and demonstrating its promising application in intelligent edge devices. As a demonstration, the proposed massively parallel computing allows the one-shot recognition of 16 letters in a neural network composed of two physically interconnected crossbars. Moreover, massively parallel computing and signal modulation are implemented simultaneously in the analogue domain, opening up unprecedented opportunities for intelligent edge applications.

Fig. 1 shows the continuous-time data representation and the corresponding continuous-time analogue computing scheme. This computing approach manifests itself in the data representation of continuous time and amplitude and takes full advantage of the physical attributes of the memristive crossbar to process data in a continuous-time domain, without suffering from issues associated with the steep rising and falling edges in digital computing. By employing Kirchhoff's current law and Ohm's law, any individual sinusoidal (or cosinusoidal) signals with different frequencies can be processed by the memristive crossbar. Moreover, we can further feed a continuous-time signal by the linear combination of sinusoidal/cosinusoidal signals into the memristive crossbar to effectively increase the computing capacity (see Fig. 1a). As shown in the middle panels of Fig. 1a, such a continuous-time signal (*e.g.,* voltage or current) in a time segment can be transformed to a frequency spectrum with multiple peaks at different frequencies. Such a transformation explicitly illustrates that the input continuous-time voltage signals of single or multiple frequency components can be processed by the memristive crossbar to output the continuous-time current signals of multiple frequency components. This transformation yields a unique process to read the information stored in the crossbar and perform computing.

It is mathematically provable to implement matrix-matrix multiplication (MMM) by using the continuous-time computing scheme in a memristive crossbar. The input voltage signal in the $i^{\text{th}}$ row ($U_i$) can be expanded by a series of orthogonal bases in the frequency domain as $U_i = \sum_k U_i^k$, where $U_i^k$ is the voltage amplitude of the $k^{\text{th}}$ frequency. Similarly, the current output in the $j^{\text{th}}$ column ($I_j$) can also be expanded as $I_j = \sum_k I_j^k$, where $I_j^k$ is the current amplitude of the $k^{\text{th}}$ frequency. By assuming a constant conductance $G_{i,j}$ during operation, the input continuous-time voltage signals are converted to the output continuous-time current signal through Ohm's law $I_{i,j} =$

$\sum_k U_i^k G_{i,j}$. The currents at all columns are summed according to Kirchhoff's current law $I_j = \sum_i I_{i,j}$. Since $I_j = \sum_k I_j^k$, we are able to achieve MMM through $I_j^k = \sum_i U_i^k G_{i,j}$ via a one-shot operation. Based on the continuous-time data representation, this frequency multiplexing computing (FMC) technology allows for the realization of massively parallel computing.

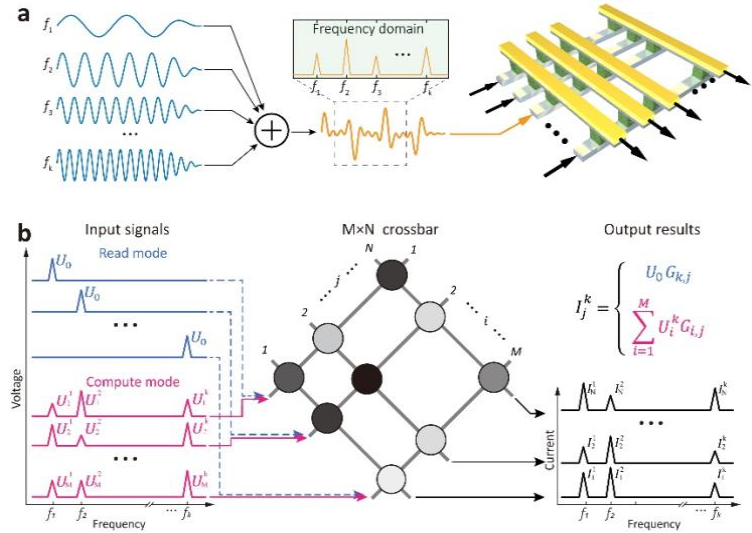

**Fig. 1 | Continuous-time data representation for frequency multiplexing computing (FMC) in a memristive crossbar. a**, Implementation of FMC by using the memristive crossbar, in which data are represented by a continuous-time signal synthesized with various sinusoidal (or cosinusoidal) signals with different frequencies $f_k$. **b**, Schematic illustration of FMC-enabled parallel reading and parallel computing. In Read mode, the parallel readout of all conductance values ($G_{k,j}$) stored in the memristive crossbar can be achieved by feeding single-frequency continuous-time signals with a constant voltage amplitude $U_0$ into a row $i$ ($i = 1, 2, ...M$) of the crossbar. In Compute mode, a one-shot MMM operation can be implemented by inputting multiple frequency continuous-time signals with the voltage amplitude $U_M^k$ at different frequency $f_k$ values into row $i$ ($i = 1, 2, ...M$) of the crossbar. For both Read and Compute modes, the output signals are generated from column $j$ ($j=1, 2,…N$) in the form of the current-frequency spectrum, in which $I_j^k$ represents the current amplitude at the frequency component $f_k$ of the $j^{th}$ column.

Implementing the one-shot MMM operation in the memristive crossbar enables massively parallel reading and computing, as schematically shown in Fig. 1b. The crossbar can be operated in either Compute mode or Read mode, which is dependent on the frequency spectrum of the input continuous-time voltage signal. When the input signal fed into each row $i$ ($i = 1, 2, ...M$) of a M × N crossbar contains a single frequency component and a constant voltage amplitude $U_0$, the parallel reading of the data stored in the crossbar is achievable. Meanwhile, when the input signal contains multiple frequency components with different voltage amplitudes, the crossbar is capable of implementing massively parallel computing. Regardless of the Read mode or the Compute mode, the output results at each column of the crossbar $j$ ($j = 1, 2, ...N$) are represented via the current-frequency spectrum.

We next implement these two FMC-based operation modes experimentally in a memristive crossbar array. Fig. 2a and 2b show scanning electron microscopy images of a fabricated Ta/HfO$_2$ memristor crossbar array and the corresponding I-V characteristics, respectively. All available resistive states in each device exhibit linear and symmetrical I-V characteristics within the voltage range from -50 mV to 50 mV. To achieve the parallel reading of the stored data in the crossbar, we applied continuous-time signals with the same voltage amplitude $U_0$ but distinct frequencies into each column of the crossbar. Subsequently, we analysed the output current-frequency spectrum and read out the conductance values by using $G_{Read} = I_{Read}/U_0$. The measured output current values (symbols) at different $U_0$ values versus pre-probed device conductance are presented in Fig. 2c. For the same $U_0$ (1, 2 or 5 mV), all measured current values are located at a straight line (dashed lines) with a slope equal to $U_0$, indicating that the accurate parallel reading is accessible in the memristive crossbar.

Sequentially, we went a step further and realized massively parallel computing. By feeding continuous-time voltage signals with 16 frequency components into all of the rows of a 25 × 9 memristive crossbar simultaneously, the corresponding continuous-time current output signals are generated instantly from all of the columns of the crossbar. We analysed the current magnitudes at 16 frequencies generated from each column in the frequency domain, with the corresponding results shown in Supplementary Fig. 1. For simplicity, we selected the experimental current values at 16 frequencies ($I_3^k$) output from the third column of the crossbar and compared them with the simulation results. Fig. 2d shows a comparison between the experimentally measured current values (grey histogram) and the simulation current values (magenta histogram) at 16 frequency components (see details for the experimental measurement and simulation in Methods).

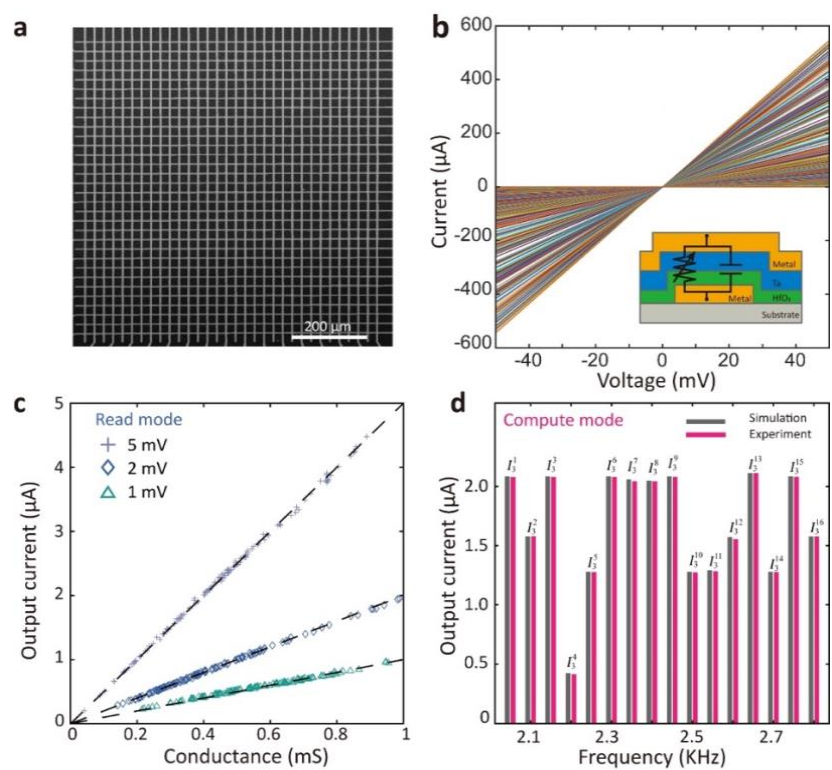

**Fig. 2 | Experimental implementations of FMC-based massively parallel computing. a,** Fabricated Ta/HfO$_2$ memristor crossbar array. **b,** Linear and symmetric I-V curves of the memristive device at different conductance states. The inset schematically shows the device structure. **c,** Measured output currents versus conductance of different devices in the memristive crossbar for $U_0 = 1, 2,$ and $5$ mV. For the same $U_0$, all the measured current values are located on a straight line with the slope equal to $U_0$, indicating that the parallel reading operation is valid. **d,** Comparison between experimental (grey histogram) and simulation current values (magenta histogram) at 16 frequencies in the massively parallel computing mode. The index labelled in each column corresponds to the output current at different frequencies.

As expected, the experimental results agree well with the simulation results, indicating the feasibility of FMC-based one-shot MMM operation. Implementing the one-shot MMM operations in the memristive crossbar will achieve the massively parallel computing of numerous tasks and allow real-time inference that is desirable for many intelligent edge applications in the IoTs network. It should be noted that the massively parallel computing proposed in this work is radically different from parallel computing based on a multicore digital processor (Supplementary Fig. 2), in which parallel computing is limited by inherently sequential computing elements and the communication bottleneck still limits the computing performance[8].

By taking advantage of FMC-based parallel reading and computing, as demonstrated above, we are capable of achieving the one-shot recognition of numerous target images by using two memristive crossbars. As shown in Fig. 3a, these two memristive crossbars are physically interconnected by trans-impedance amplifiers (TIAs). The left crossbar in this prototype is used to store target letter images shown in Fig. 3b, and the right crossbar is used as an artificial neural network for inference. To demonstrate the one-shot recognition of numerous target images, we mapped 16 letter images "NAINVINAJNGINUHC" corresponding to a $25 \times 16$ data matrix (Fig. 3b) into the left crossbar and the trained weight matrix (Fig. 3c) into the right crossbar. Subsequently, 16 different carrier signals, which have the same voltage amplitude $U_0$ but different frequencies (*i.e.,* from $f_1$ to $f_{16}$), were simultaneously fed to all columns of the left crossbar to carry out parallel reading (in FMC-based Read mode). The output current signals from the left crossbar that represent the 16 target letters are converted into continuous-time voltage signals and then input into the right crossbar for classifying the target letters (in FMC-based Compute mode). The recognition results are output from the right crossbar. With this unique setup, the 16 letter images can be classified into 9 different categories in a massively parallel one-shot manner, which are labelled 'A', 'N', 'J', 'I', 'G', 'V', 'U', 'C' and 'H'.

Note that signal modulation can be accomplished simultaneously with the one-shot recognition of numerous target letters in this FMC-based system, which is different from digital technologies where the signal modulation and image processing are separated. In this way, the recognition results output from the right crossbar can be directly transmitted through wireless channels (Tx_1, Tx_2… Tx_9) by using a radio frequency (RF) module. The transmitted signals can be received on a remote terminal device over different wireless channels (Rx_1, Rx_2… Rx_9), as shown in Fig. 3d, in which the red boxes represent the classified target letters. We also experimentally demonstrated the reception of the recognition results on a remote mobile phone (Supplementary Fig. 3). With FMC-based technology, we are able to achieve massively parallel one-shot recognition of numerous target images, as well as signal modulation, transmission and reception in real time. This transformative technology is desirable for advancing intelligent edge devices in IoT networks that demand high-efficiency data processing and communication. Moreover, we demonstrate that the FMC-based system is compatible with multiple-input-multiple-output (MIMO) communication technology widely used for increasing the channel capacity of 5G wireless communication networks[40] (Supplementary Fig. 4).

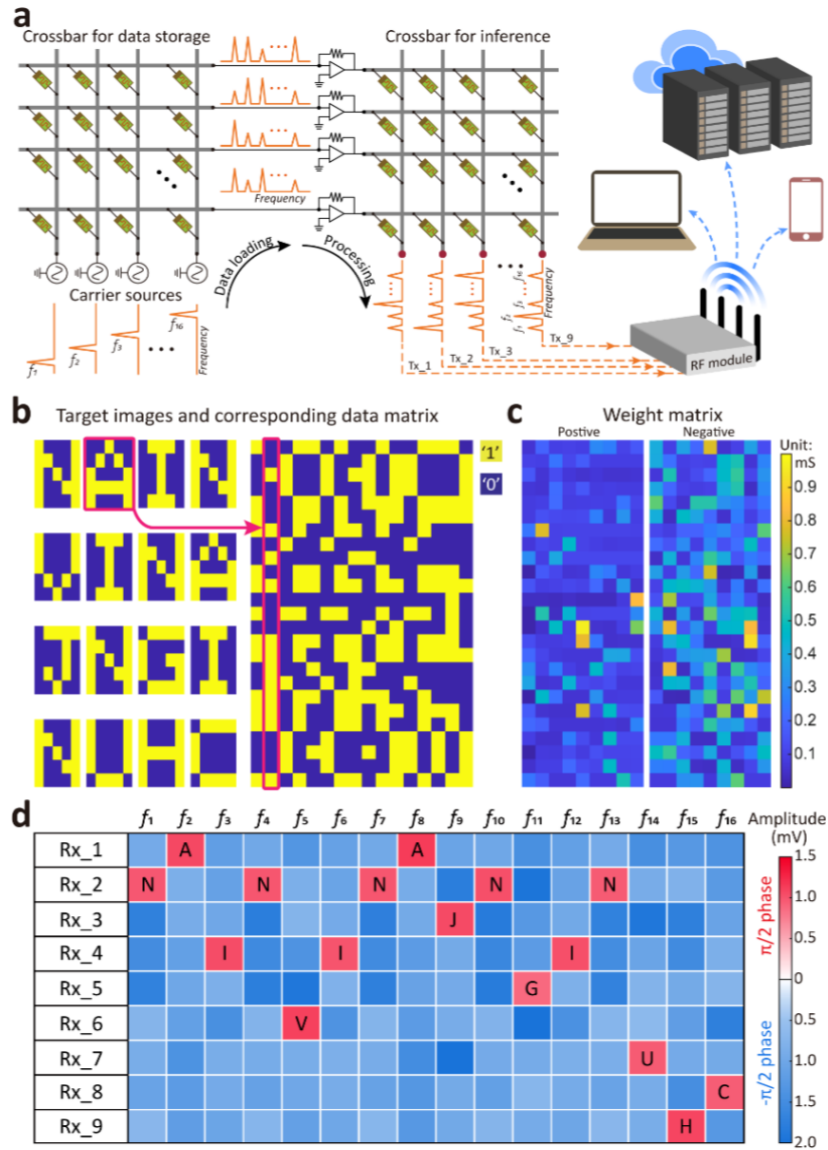

**Fig. 3 | FMC-based one-shot recognition of numerous images and wireless communication of the recognition results. a**, The circuit schematic of the device based on two crossbars and an RF module. The left and right crossbars are used for data storage and inference, respectively. **b**, The data matrix for 16 target images "NAINVINAJNGINUHC" stored in the left crossbar. **c**, The trained weight matrix for the right crossbar. **d**, One-shot recognition results of the 16 letter images. The recognition results were transmitted via the RF module and received by the terminal device through wireless communications. The red boxes in each row (Rx_1 to Rx_9) correspond to the classified target letters from the 16 input letters represented by different frequency components ($f_1, f_2 \cdots f_{16}$).

The use of the continuous-time data representation offers tremendous promise for reducing the operating voltage and increasing the computing frequency. Since the voltage $U_0$ of the carrier signal input into the left crossbar in Fig. 3a is critical to the accuracy of the FMC-based massively parallel computing, we evaluated the parallel reading errors by using $(G_r - G_R)/G_r$ at different $U_0$ values, *i.e.,* from 1 mV to 100 mV (Fig. 4a and Supplementary Fig. 5), where $G_r$ represents the readout conductance of the memristive crossbar by the semiconductor parameter analyser and $G_R$ represents the conductance values obtained by the FMC. Our results show that the relative error is less than 2 % at $U_0 = 2$ mV and is comparable to that reported in neuromorphic computing[37,39,41]. Different from the small error at low operating voltage, high operating voltage induced harmonic distortion and conductance variation would lead to a large error (Supplementary Fig. 6). Note that a low-precision computation is sufficient for most neural network applications[29,36]. For similar-level precision, it is highly desirable to make the neural networks operate at a low voltage to achieve high energy efficiency. The operating voltage of the neural network based on the continuous-time data representation is two orders of magnitude lower than that of the digital circuit-based neural network[42-44] and reported values for the neural networks implemented with the memristive crossbar[35,38,41,45-47]. We reveal that the reason why massively parallel computing can be operated at ultralow voltage is due to highly suppressed noise at a high operating frequency (Fig. 4b). As the frequency increases, the signal-to-noise ratio (SNR) is improved (inset of Fig. 4b). Although the proof-of-concept is demonstrated at a frequency of a few kHz, it should be pointed out that the operating frequency can be further increased. To explore the upper limit of the operating frequency, we carried out scattering-parameter (S-parameter) measurements for the memristive devices (0.4×0.4 µm²), with the results shown in Fig. 4c. The experimental S21 measurement matches well with the simulation model based on the equivalent circuit of the memristive devices (Supplementary Fig. 7), from which the capacitance can be extracted. Based on the extracted capacitance (~20 fF), we obtained an operating frequency of 5 GHz even for a memristive crossbar array made of the large-area memristive device (see Methods for more details). By reducing the feature size of the memristive device[48,49], it is possible to engineer the operating frequency of the FMC-based massively parallel computing well beyond the clock frequency available for digital computing (Fig. 4d). Such a wide operating frequency range enables to add more frequency components in the input continuous-time signal to increase the parallel computing capability. In conjunction with recent advance in the integration density of the memristive crossbar[33,38], the parallel processing capability would be dramatically enhanced.

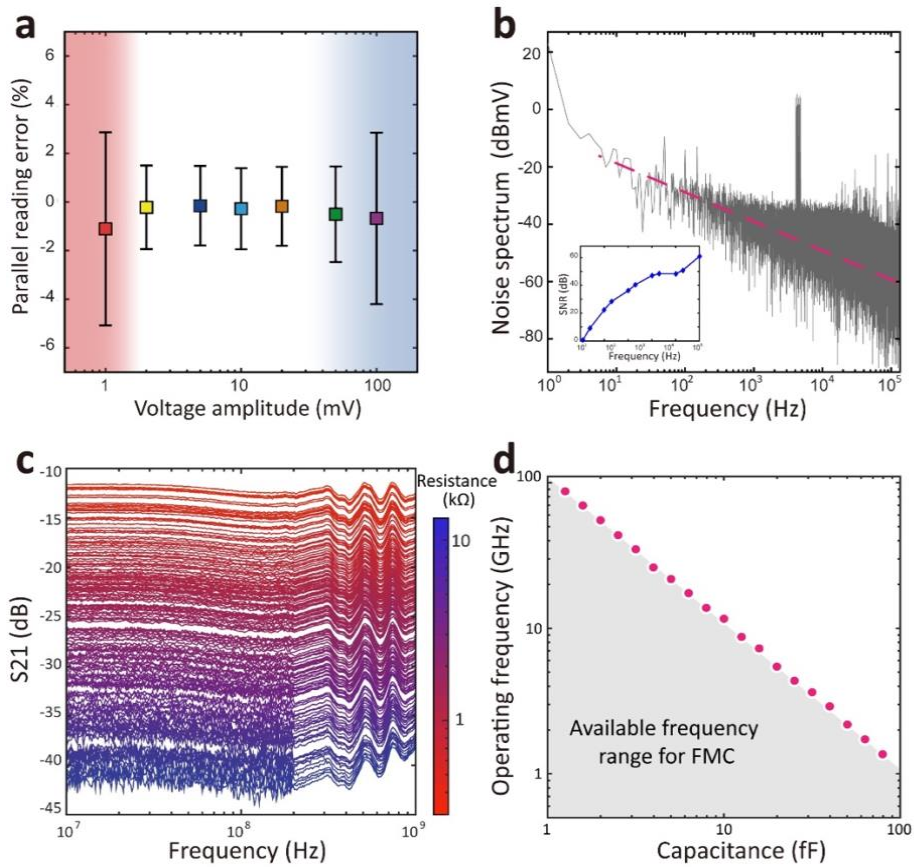

**Fig. 4 | Performance of FMC-based massively parallel computing. a**, The relative error is evaluated at various voltage amplitudes ($U_0$) of the carrier signals. **b**, Noise spectrum measured for the memristive device. The dashed red line is used to indicate the trend of noise suppression with increasing operating frequency. **c**, Scattering-parameter S21 measurement of the memristive device by using a vector network analyser at different conductance levels (solid lines). **d**, The operating frequency of the FMC-based massively parallel computing is calculated based on the capacitance of the memristive device.

In conclusion, we propose and experimentally demonstrate a frequency multiplexing-enabled massively parallel computing scheme based on a continuous-time data representation. The ultralow operating voltage and ultrahigh operating frequency of the FMC-based massively parallel computing may open up an avenue for parallel analogue computers that are superior to their digital counterparts with orders of magnitude improvement in the voltage and computing speed. In conjunction with the signal modulation accomplished simultaneously with parallel signal processing, FMC-based massively parallel computing may be deployed for low-power intelligent edge applications to address the upcoming challenges associated with real-time processing and communication in IoT networks[50]. The proposed FMC could be extended to other crossbars made of non-volatile memory devices[13,32,51-56] (*e.g.,* phase change memory and magnetoresistive random access memory).

## Methods

### Device fabrication

Pd/Ta/HfO$_2$/Pd memristive devices were fabricated with a sandwich structure. An Al$_2$O$_3$ substrate was used to eliminate the parasitic capacitance. The bottom/top metal layer was deposited through a standard electron beam deposition process; a 6-nm-thick HfO$_2$ switching layer was deposited via atomic layer deposition. The deposition of an 80-nm-thick Ta layer was realized by using a standard RF sputtering process. For devices with a feature linewidth larger than 2 μm, the electrode patterns were realized by using double-layer photoresist photolithography, followed by a lift-off process in N-methyl pyrrolidone. For linewidths smaller than 2 μm, electron beam lithography was used to pattern the electrodes.

### Implementation of massively parallel computing in the memristive crossbar

To demonstrate massively parallel computing in the memristive crossbar array, signal generators, TIAs and an oscilloscope with a built-in frequency analyser were used. Signal generators supporting 16 channels were used as continuous-time signal sources. We employed TIAs to convert the current signals into voltage signals for the measurement; a frequency analyser was used to measure the output results from the memristive crossbar array.

**Implementation of parallel reading in the memristive crossbar**

To demonstrate the parallel reading of the data stored in the memory crossbar, we used the individual-frequency sinusoidal voltage signal as carrier signals and applied them into different rows of the crossbar, where the conductance matrix is written randomly. The output current was converted into a voltage by the TIAs and was subsequently measured by the frequency analyser. In Read mode, multiple current peaks are present in the current-frequency spectrum output from each column of the crossbar. Based on the specific row in which a carrier signal with distinct frequency was applied, the frequency corresponding to a specific current peak can be distinguished at each column of the crossbar, in which the conductance of the selected memristor is proportional to the specific current magnitude (or peak).

**Measurement of the S21 parameter on the memristive device**

We used a Tektronix TTR506A vector network analyser to apply high-frequency microwave signals to the memristive devices to measure the S-parameter. The S21 parameter represents the forward transmission gain through the memristive devices. The S21 curves in Fig. 4c were measured on a memristive device with an area of 0.16 $\mu m^2$. Different S21 curves were obtained at different conductance levels (ranging from 300 to 13000 Ω). A simulation model with a variable resistor connected in parallel with a constant capacitor was developed to calculate S21. The simulation results are in good agreement with the experimental results (see Supplementary Fig. 7). By fitting to the measured S21 curves with the model, we extracted the capacitance of the 0.16 $\mu m^2$ memristor to be 20 fF.

**Operating frequency of FMC-based massively parallel computing scheme**

To explore the full potential of massively parallel computing, we analysed the operating frequency of FMC-based massively parallel computing implemented in a memristive crossbar array. For simplicity, a memristor has been considered a programmable resistor connected in parallel with a parasitic capacitance. In the simulation, we used a differential pair of memristive devices as synaptic weight. Since the leakage current through parasitic capacitor increases with the frequency, the maximum operating frequency available for FMC-based massively parallel computing is thus limited by the capacitance variation in different memristive devices. To carry out the simulation, we used a 512×512 memristive crossbar array, in which capacitance variation follows a normal distribution with a relative standard deviation of 10% and memristor resistance varies from 1 to 10 kΩ. Within 1% error in massively parallel computing, the operating frequency was calculated versus capacitance of memristive device and shown in Fig. 4d.

**Training of the artificial neural network**

To reduce the power consumption in the memristor arrays and mitigate the effect of the wire resistance, high resistance states were preferred when programming the memristive crossbar. A weight change was realized by stimulating an alternative memristor in a differential pair, $W=G_+-G_-$. All memristors in both positive and negative crossbar arrays were first set into high resistive states. Based on the gradient descent algorithm, an online training process was performed. Subsequently, the expected conductance was mapped into a memristive crossbar. If $\Delta G(i,j)>0$, then a positive pulse would be applied to increase $G_+$, and a negative pulse would be used to decrease $G_-$; Otherwise, a positive pulse would be applied to increase $G_-$, and a negative pulse would be used to decrease $G_+$.

**Channel capacity of the MIMO-OFDM communication system**

A wide-band MIMO-orthogonal frequency-division multiplexing (OFDM) communication system with $N_t$ transmitting antennas and $N_r$ receiving antennas was considered. The MIMO system model at the $k^{th}$ subcarrier is given as

$$y_k = H_k s_k + n_k, \tag{1}$$

where $y_k \in \mathbb{C}^{N_r}$ and $s_k \in \Omega^{N_t}$ are the received and transmitted vectors at the $k^{th}$ subcarrier, respectively, and the constellation of each component $s_i$ is denoted by $\Omega$. $H_k \in \mathbb{C}^{N_r \times N_t}$ is the channel matrix at the $k^{th}$ subcarrier and assumed to be perfectly known at the receiving terminal, and $n_k \sim \mathcal{CN}(0, \sigma_k^2 I_{N_r})$ is the additive white Gaussian noise (AWGN) vector at the $k^{th}$ subcarrier. By applying information theory to

the MIMO-OFDM system model, the capacity can be obtained as follows:

$$C = \sum_k B_k \max_{S_k: tr(S_k) \leq P_k} \log_2 \left| I_{N_r} + \frac{1}{\sigma_k^2} H_k S_k H_k^H \right|, \qquad (2)$$

where $C$ is the channel capacity of the MIMO-OFDM system, $B_k$ and $P_k$ are the bandwidth and transmitting power of the $k$th subcarrier, respectively, and $S_k$ denotes the covariance matrix of the transmitting signal vector at the $k^{th}$ subcarrier. Note that the bandwidth of each subcarrier is set to be the same in a realistic OFDM system. The average channel capacity with respect to the number of transmitting/receiving antennas can be evaluated for different numbers of subcarriers. Supplementary Fig. 4 shows the simulation results, in which an independent and identically distributed Rayleigh-fading channel matrix was used. As shown in the figure, the channel capacity of the MIMO-OFDM system is approximately proportional to the number of transmitting/receiving antennas at a large-scale antenna array. In addition, the channel capacity can also be enhanced by increasing the number of subcarriers. These results indicate that the transmission rate of the recognized results output from the FMC system can be further increased by adopting MIMO technology and increasing the number of subcarriers.

**Acknowledgements**

This work was supported in part by the National Natural Science Foundation of China (61625402, 62034004, 61921005, 61974176), and the Collaborative Innovation Center of Advanced Microstructures and Natural Science Foundation of Jiangsu Province (BK20180330), Fundamental Research Funds for the Central Universities (020414380084).

**Author Contributions**

F.M., S.J.L. and C.W. conceived the idea and designed the experiments. F.M. and S.J.L. supervised the whole project. C.W. performed all experiments. C.W. and S.J.L. analysed experimental data. C.Y.W. and P.C. provided assistance during experiment design. Z.Z.Y. assisted in the device fabrication and circuit assembly. X.S. and W.W. contributed to circuit measurement. Y. G., Z.Z. and C. Z. contributed to MIMO model. C.W., S.J.L. and F.M. co-wrote the manuscript.

## Competing interests

The authors declare no competing interests.

## Data availability.

The data supporting the findings of this study are available within the article and its Supplementary Information, and from the corresponding authors upon reasonable request.